%% file: main.tex
\documentclass{article}
\usepackage{icmc2025_paper_template}
\usepackage{times}
\usepackage{mathptmx}
\usepackage{ifpdf}
\usepackage{soul}
\usepackage[english]{babel}
\usepackage{subfiles}
\usepackage{numprint}
\usepackage[pdftex]{graphicx}
\usepackage{tikz}
\usetikzlibrary{shapes.geometric, arrows, positioning}
\usepackage{siunitx}
\usepackage{float}
\usepackage{booktabs}
\usepackage{microtype}  % optischer Randausgleich etc.
\usepackage{amsmath}
\usepackage{svg}

\def\note[#1#2#3]{#1\if b#2$\flat_#3$\else\if#2##$\sharp_#3$\else$_#2$\fi\fi}

\def\papertitle{Fretting-Transformer: Encoder-Decoder Model for MIDI to Tablature Transcription}
\def\firstauthor{Anna Hamberger}
\def\secondauthor{Sebastian Murgul}
\def\thirdauthor{Jochen Schmidt}
\def\fourthauthor{Michael Heizmann}
\newif\ifpdf
\ifx\pdfoutput\relax
\else
   \ifcase\pdfoutput
      \pdffalse
   \else
      \pdftrue
  \fi
\fi

\ifpdf % compiling with pdflatex
  \usepackage[pdftex,
    pdftitle={\papertitle},
    pdfauthor={\firstauthor, \secondauthor, \thirdauthor},
    bookmarksnumbered, % use section numbers with bookmarks
    pdfstartview=XYZ % start with zoom=100% instead of full screen; 
   ]{hyperref}
  \graphicspath{{./figures/}}
  \DeclareGraphicsExtensions{.pdf,.jpeg,.png}

  \usepackage[figure,table]{hypcap}

\else % compiling with latex
  \usepackage[dvips,
    bookmarksnumbered, % use section numbers with bookmarks
    pdfstartview=XYZ % start with zoom=100% instead of full screen
  ]{hyperref}  % hyperrefs are active in the pdf file after conversion

  \usepackage[dvips]{epsfig,graphicx}
  \graphicspath{{./figures/}}
  \DeclareGraphicsExtensions{.eps}

  \usepackage[figure,table]{hypcap}
\fi

\hypersetup{
    colorlinks,%
    citecolor=black,%
    filecolor=black,%
    linkcolor=black,%
    urlcolor=black
}

\title{\papertitle}

\fourauthors
  {0.6in}
  {\firstauthor$^*$} {Rosenheim Technical University of Applied Sciences \\ %
    {\tt \href{mailto:anna-hamberger@web.de}{anna-hamberger@web.de}}}
  {\secondauthor$^*$} {Klangio GmbH \\ %
    {\tt \href{mailto:sebastian.murgul@klang.io}{sebastian.murgul@klang.io}}}
  {\thirdauthor} { Rosenheim Technical University of Applied Sciences \\ %
    {\tt \href{mailto:jochen.schmidt@th-rosenheim.de}{jochen.schmidt@th-rosenheim.de}}}
  {\fourthauthor} {  Karlsruhe Institute of Technology \\ %
    {\tt \href{mailto:michael.heizmann@kit.edu}{michael.heizmann@kit.edu}}}

\begin{document}
\capstartfalse
\maketitle
\capstarttrue
{\def\thefootnote{*}\footnotetext{These authors contributed equally to this work.}}
\begin{abstract}
\subfile{sections/abstract}
\end{abstract}

\section{Introduction}\label{sec:introduction}
\subfile{sections/introduction}

\section{Related Work}\label{sec:related-work}
\subfile{sections/related_work}

\section{Methodology}\label{sec:methodology}
\subfile{sections/methodology}

\section{Experiments and Results}\label{sec:experiments}
\subfile{sections/experiments}

% \pagebreak

\section{Conclusions}
\subfile{sections/conclusion}

% \begin{acknowledgments}
% At the end of the Conclusions, acknowledgments to people, projects, funding agencies, etc. can be included after the second-level heading ``Acknowledgments'' (with no numbering).
% \end{acknowledgments} 

%%%%%%%%%%%%%%%%%%%%%%%%%%%%%%%%%%%%%%%%%%%%%%%%%%%%%%%%%%%%%%%%%%%%%%%%%%%%%
%bibliography here
\bibliography{references}

\end{document}

%% file: sections/abstract.tex
%%% abstract
Music transcription plays a pivotal role in Music Information Retrieval (MIR), particularly for stringed instruments like the guitar, where symbolic music notations such as MIDI lack crucial playability information. This contribution introduces the Fretting-Transformer, an encoder-decoder model that utilizes a T5 transformer architecture to automate the transcription of MIDI sequences into guitar tablature. By framing the task as a symbolic translation problem, the model addresses key challenges, including string-fret ambiguity and physical playability. The proposed system leverages diverse datasets, including DadaGP, GuitarToday, and Leduc, with novel data pre-processing and tokenization strategies. We have developed metrics for tablature accuracy and playability to quantitatively evaluate the performance. The experimental results demonstrate that the Fretting-Transformer surpasses baseline methods like A$^*$ and commercial applications like Guitar Pro. The integration of context-sensitive processing and tuning/capo conditioning further enhances the model's performance, laying a robust foundation for future developments in automated guitar transcription.

%% file: sections/introduction.tex
%%% introduction
Notation-level music transcription is the process of converting musical audio or symbolic data into a written form. This task is both challenging and essential in the field of Music Information Retrieval (MIR) \cite{Benetos.2019}. Automatic Music Transcription (AMT) seeks to address the limitations of manual transcription by creating algorithms that can transform musical input into symbolic representations. 

For guitarists, the use of classical Western music notation is rather unusual because it does not contain specific information about string choice and fret position, which are essential for playing the instrument, since the same pitch can be played in different ways on the guitar. Here, tablature (tab) notation is preferred, which is optimized for stringed and fretted instruments and provides a visual representation of the placement of the fingers on the guitar fretboard. However, it is a challenging task to convert a piece of music from symbolic music notation such as MIDI to tablature, as it requires a deep understanding of both music theory and the characteristics of guitar playing. It is not just a matter of choosing the right strings and frets in terms of pitch, but also to ensure that these frets are physically playable and hand forms are incorporated that the player is used to. 

In this research, we tackle the challenge of transcribing MIDI data into acoustic guitar tablatures. This task presents several difficulties, including encoding pitches as well as musical context, resolving ambiguities in fretboard positions, and ensuring realistic playability. Although most existing methods (see Section \ref{sec:related-work}) rely on manually defined objective functions, e.\,g., based on minimizing finger stretching or movement between positions, this work leverages a state-of-the-art T5 transformer model \cite{Raffel.2019}. By treating guitar transcription as a symbolic translation problem, the approach produces accurate and playable tablatures, addressing both musical and physical constraints.

%% file: sections/related_work.tex
%%% state of the art
There are various approaches in the field of automatic guitar tablature transcription, including rule-based, probabilistic, graph-based, and neural network-based methods.
% rulebased
Early systems relied on predefined rules to create tablatures from MIDI data. The software developed by Wang and Li utilizes harmonic rules and fretting styles to produce scores, but often requires manual adjustments to ensure they were playable \cite{Wang.1997}. 
Miura et al.\ enhanced these methods by minimizing hand movements; however, advanced players found these limitations to be too restrictive \cite{Miura.2004}.

% Probabilistic
Genetic algorithms (GAs) are employed to optimize tablatures based on playability criteria. Tuohy and Potter \cite{Tuohy.2005} introduced a GA designed to generate playable fret positions. Ramos et al.\ \cite{Ramos.2015} built upon this by enhancing the algorithm with subpopulation techniques, which improved its efficiency. More recently, Bastas et al.\ \cite{Bastas.2022} have integrated string-related audio features into a GA, resulting in more refined outcomes.
Hidden Markov Models (HMMs) have also been applied, where states represent string configurations and the transitions are influenced by physical difficulties. Barbancho et al.\ \cite{Barbancho.2012} used the Viterbi algorithm to map audio signals to optimal finger positions, achieving more accurate results.

% graph-based
Graph-based techniques represent the relationships between notes, strings, and frets as a directed acyclic graph. In 1989, Sayegh \cite{Sayegh.1989} introduced the optimal path paradigm, which assigns transition costs based on hand movements. Subsequent extensions by Radicioni et al.\ incorporated biomechanical constraints and hand span considerations \cite{Radicioni.2004}. 
% A-Star
Burlet and Fujinaga \cite{Burlet.2013} have built upon Sayegh's approach by developing a new algorithm for guitar tablature transcription called A-star-Guitar. This algorithm utilizes the A$^*$ pathfinding method to create optimal guitar tablatures for polyphonic music. It works by searching for the optimal path in a graph that includes all possible combinations of strings and frets for notes and chords. The algorithm takes into account the tuning of the guitar, the number of frets, and the position of the capo. In this graph, possible fretboard positions are represented as nodes, while the edges are weighted based on biomechanical factors, such as the difficulty in moving between frets, the finger span required for chords, and penalties for positions beyond the seventh fret. The algorithm employs a heuristic function that calculates the cumulative weight of edges from a given vertex to the target vertex, enabling it to identify the easiest transitions between notes.

% neuronal network based
Neural networks, especially convolutional neural networks (CNN), have shown significant potential in tablature transcription. Wiggins and Kim introduced TabCNN, which maps spectrogram images to tablatures \cite{Wiggins.2019}. Kim et al.\ improved the approach by integrating self-attention mechanisms, resulting in improved transcription accuracy and better long-term sequence modeling \cite{Kim.2022}.
% transformer in music generation
Recent advances in deep learning have established transformers as a leading architecture for analyzing sequential data, including music generation. Transformers are particularly effective in capturing long-term dependencies and complex patterns, enabling them to generate music with both temporal and harmonic consistency. 
Early work, such as the Music Transformer, introduced relative positional encoding to better address the nuances of pitch and timing in music \cite{Huang.2018}. Subsequent models, including the Pop Music Transformer \cite{Huang.2020} and Theme Transformer \cite{Shih.2023}, further refined these techniques, focusing on rhythmic structure and thematic consistency. 
In 2021, Sarmento et al.\ \cite{Sarmento.2021} created a diverse symbolic dataset called DadaGP, which contains \numprint{26181} files representing 739 music genres. They also developed a token format based on event-based MIDI encoding. To evaluate the dataset and the token format, the authors trained a Pop Music Transformer \cite{Huang.2020} to generate new symbolic compositions.
In their work, Chen et al.\ \cite{Chen.2020} adapted Transformer XL \cite{Dai.2019} specifically to generate fingerstyle guitar tablatures. They enhanced the model by incorporating tokens for string and fret positions, in addition to pitch and duration. While their approach successfully produced valid tablatures, it faced challenges with string assignments for lower pitches and did not adequately evaluate long-term musical structure. These issues highlight the need for further refinement in the application of transformers for guitar-related tasks.
% BERT Quelle
Recent advances have introduced transformer models for transcribing MIDI into tablature. Edwards et al.\ \cite{Edwards.2024} utilize the BERT model \cite{Devlin.2019}, which they train by tokenizing MIDI data using the MidiTok method \cite{Fradet.2023} and masking the string tokens within the input sequence. 
Initially, they trained the model on the entire DadaGP dataset \cite{Sarmento.2021}, followed by fine-tuning on a selected set of professional transcriptions from the Leduc dataset presented in \cite{Riley.2024.High}. An evaluation study involving guitarists demonstrated that the transformer model outperformed other methods in terms of playability. Although this approach is quite promising, it is limited to standard guitar tuning and does not allow the use of a capo.

%% outro
Transformer models have mainly been used for music generation and only recently for guitar tablature transcription. However, their ability to learn musical structures makes them a promising tool for this task. Although several probabilistic and graph-based approaches have been explored, there are only limited studies focusing on MIDI-based transcription using neural networks. Therefore, this research aims to investigate the potential of transformer-based approaches further.

%% file: sections/methodology.tex
% Datasets
\subsection{Datasets}
\label{s:datasets}
Unlike datasets that concentrate on converting audio into symbolic formats through complex Automatic Music Transcription (AMT) pipelines \cite{Benetos.2019}, the datasets used in this research – DadaGP, GuitarToday and Leduc – focus on symbolic data provided in the Guitar Pro\footnote{\url{https://www.guitar-pro.com/}} format. This format allows for direct experimentation with MIDI tablature transcription by encoding pitch as well as string/fret information. The following paragraphs summarize the characteristics and relevance of these datasets.

The GuitarToday dataset\footnote{\url{https://www.fingerstyle-guitar-today.com}} contains $363$ easy fingerstyle guitar tablatures designed for beginners and intermediate players. Sourced from the `GuitarToday' Patreon account\footnote{\url{https://www.patreon.com/guitartoday}}, the dataset features tracks in standard tuning (\note[E4], \note[B3], \note[G3], \note[D3], \note[A2], \note[E2]) and focuses on simpler pieces. The dataset analysis shows a predominance of beginner-friendly pitches with open strings or low fret positions. These characteristics make it an ideal foundation for model training with minimal complexity.

The DadaGP dataset contains over \numprint{26000} tracks across various genres, including rock, metal and classical music \cite{Sarmento.2021}. After filtering, a total of \numprint{2301} acoustic guitar tracks were selected for this study. This dataset includes a range of note durations and a broader pitch spectrum compared to GuitarToday, reflecting more complex musical compositions. Most tracks are in standard tuning, although there are occasional variations like drop tunings and the use of capos. The distribution of string-fret combinations is wider, with a notable emphasis on mid-range frets. The DadaGP dataset was compiled from Ultimate Guitar\footnote{\url{https://www.ultimate-guitar.com/}}, a platform where the quality of contributions varies greatly. This variability requires caution when interpreting the results derived from the DadaGP dataset. However, it serves as a valuable complement to GuitarToday, introducing the model to a greater diversity of musical and technical contexts.

The Leduc dataset consists of $232$ jazz guitar tablatures created by François Leduc \cite{leduc.2024}. It highlights the rich harmonic and rhythmic complexity characteristic of jazz music. Although the dataset is relatively small, it provides valuable insights into jazz-specific playing styles, including mid-range pitch preferences and intricate chord voicings. These features make it a useful addition to other datasets, enhancing the model's ability to generalize across different musical genres.

% Data Pre-Processing
\subsection{Data Pre-Processing}
\label{s:pre-processing}
\begin{figure}[tb]
    \footnotesize
    \centering
    
    \begin{tikzpicture}[
        auto,
        node distance=0.5cm, 
        roundnode/.append style={ellipse, draw=red!60, fill=red!5, very thick, minimum width = 2cm, minimum height = 1.2cm, text width=4.5em, align=center},
        process/.append style={rectangle, rounded corners, minimum width=2cm, minimum height=1cm, align=center, draw=black, fill=black!5, inner sep=5pt}, 
        arrow/.style={thick,->,>=stealth}]

        % Nodes
        \node (gp-file)[roundnode] {Guitar Pro-File};
    
        \node (filter) [process, right = of gp-file] {Filter Acoustic \\ Guitar Tracks};
        \node (data) [process, right = of filter] {Data \\ Clean Up};
        \node (duplicates) [process, below = of data] {Remove \\ Duplicates};
        \node (split) [process, left = of duplicates] {Train/Valid/Test \\ Split};
        \node (midi) [process, left = of split] {Convert to \\ MIDI Files};
    
        \node (midi-file)[roundnode, below = of midi] {MIDI-File};
    
        \node (extract) [process, right = of midi-file] {Extract Relevant \\ Information};
        \node (encoding) [process, right = of extract] {Encoding and \\ Tokenization};
        \node (sequence) [process, below = of encoding] {Split into \\ Sequences};
        \node (dataset)[roundnode, left = of sequence] {Dataset};
    
        % Arrows
        \draw [arrow] (gp-file) -- (filter);
        \draw [arrow] (filter) -- (data);
        \draw [arrow] (data) -- (duplicates);
        \draw [arrow] (duplicates) -- (split);
        \draw [arrow] (split) -- (midi);
        \draw [arrow] (midi) -- (midi-file);
        \draw [arrow] (midi-file) -- (extract);
        \draw [arrow] (extract) -- (encoding);
        \draw [arrow] (encoding) -- (sequence);
        \draw [arrow] (sequence) -- (dataset);

    \end{tikzpicture}
    \caption{Flowchart of the data pre-processing steps}
    \label{f:pre-processing}
\end{figure}
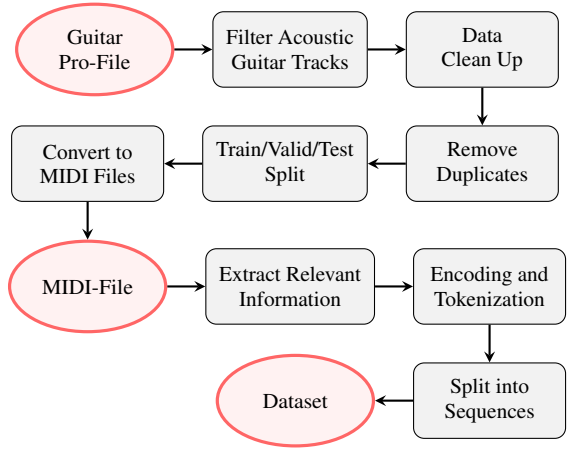

To train transformer models for guitar tablature transcription, Guitar Pro files are pre-processed to extract relevant MIDI information and convert it into a text-based format suitable for encoding. This process involves several stages, as illustrated in Figure \ref{f:pre-processing}.

The GuitarToday and Leduc datasets consist of single-track acoustic guitar files that require minimal cleanup. In contrast, the DadaGP dataset includes arrangements for multiple instruments and needs filtering to isolate the acoustic guitar tracks. This process is accomplished by utilizing MIDI channel IDs, specifically 25 for nylon stringed guitars and 26 for steel stringed guitars. Due to inconsistencies in the assignment of instrument IDs, additional keyword-based filtering is applied using guitar-related terms in various languages to ensure that we focus only on the relevant tracks. As a result, about $5\%$ of \numprint{47039} tracks could be used.
Duplicate tracks are removed to ensure dataset uniqueness by matching metadata or file names. The examples are split into training, validation, and test sets (90/10/10), maintaining diversity in key, tuning, and capo usage.
GuitarPro files are converted to MIDI format using the Python packages PyGuitarPro\footnote{\url{https://pyguitarpro.readthedocs.io}} and mido\footnote{\url{https://mido.readthedocs.io}}. The hierarchical structure of Guitar Pro, which includes tracks, bars, voices, beats, and notes, is simplified when converting to MIDI, as MIDI represents music as a sequence of messages.
The relevant attributes -- start time, end time, pitch, string, and fret -- are extracted from MIDI files to represent the musical content in text format. Data tokenization divides the extracted MIDI information into sequences for model training. Five different encoding schemes were developed to explore different levels of abstraction and granularity (see \ref{s:data-encodings}). Word-level tokenizers assign numerical IDs to the tokens, resulting in datasets of varying sequence lengths for experimentation. This encoding approach enables a systematic analysis of how data representation impacts model performance.

\subsection{Data Augmentation}
\label{s:data-augmentation}
For training purposes, the three datasets are combined. Due to the significant imbalance in the dataset regarding the different capo uses and tunings, we extended the dataset to develop a conditioned model. Since the capo condition affects all pitches simultaneously while keeping the string-fret combinations unchanged, augmenting the dataset is straightforward. To augment the capo usage, we first filtered the dataset to include only files in standard tuning. Then, we artificially transformed each file from capo zero to capo seven, ensuring that each piece still contained valid string and fret combinations for the given capo. To reduce the size of the test dataset to 150 files, since not every capo variant is necessary for every piece of music, we systematically iterated through the test files and applied the next capo number where applicable. This approach ensures that every piece of music is represented while considering different variants of the capo. The tuning augmentation was applied to each training sequence. A tuning was randomly selected for each sequence and the pitches of the notes were adjusted accordingly. The four most common tunings standard, half-step down, full-step down, and drop-d were utilized.

% Model (Training)
\subsection{Model}
\label{s:model}
In our research, we define the task of transcribing pitches into tablature as a translation problem, where the model learns to map MIDI note sequences to their corresponding guitar tablatures. With this focus on the translation paradigm, we have selected the T5 model as an ideal candidate for training and fine-tuning this specific task.
The T5 architecture, which stands for Text-to-Text Transfer Transformer, was introduced by Raffel et al.\ in 2020 \cite{Raffel.2019}. The T5 represents a significant advance in natural language processing (NLP) by framing all tasks as text-to-text problems. This means that both inputs and outputs are treated as text. 

The Hugging Face Transformers package is used for implementing the network. We employ a reduced architecture of the T5 model, halving the configuration of \textit{t5-small} with a model dimension $d_\textit{model}=128$, feedforward dimension $d_\textit{ff}=1024$, three encoder-decoder layers and four attention heads. This model is trained from scratch, utilizing the Adafactor optimizer with a self-adaptive learning rate. To tokenize the data, we found that an event-based approach is optimal, similar to the encoding used in the Music Transformer \cite{Huang.2018}. The input consists of \textit{NOTE\_ON} and \textit{NOTE\_OFF} events, along with \textit{TIME\_SHIFT} tokens for timing. The output uses \textit{TAB$<$\#,\#$>$} tokens, which represent both the string and fret numbers, followed again by the \textit{TIME\_SHIFT} token. In total, we trained two versions of the model using the three combined datasets. The standard model is based on standard guitar configurations to test general functionality. The conditioned model contains a \textit{CAPO$<$\#$>$} and \textit{TUNING$<$\#,\#,\#,\#,\#,\#$>$} token in the input to condition the model for more flexibility and control over the outputs. Both models were trained on input sequences of 512 tokens length. In total, the standard dataset included \numprint{16451} training sequences and \numprint{1819} validation sequences, while the conditioned dataset comprised \numprint{129748} training sequences and \numprint{14365} validation sequences. During inference, chunks of 20 notes are processed. The tokens from the last note of the previous chunk are placed at the beginning of the following sequence in both the encoder and decoder to preserve context between the chunks.

% Data Post-Processing (Corrections)
\subsection{Data Post-Processing}
\label{s:post-processing}
In some cases, the model can generate tabs for a note that results in an incorrect pitch. To address this, errors are corrected in a post-processing step to ensure that the piece of music remains unchanged. Our post-processing algorithm refines the model's output by comparing the estimated note sequence to the corresponding input note sequence. It attempts to match each input note to its closest counterpart in the estimated sequence within a configurable window of $\pm5$ notes. The algorithm evaluates pitch values and selects the best match. If no direct match is found, the first viable string-fret combination generated for the guitar configuration used is applied. That way, we ensure that the tablatures reflect the original notes.

% Evaluation Metrics
\subsection{Evaluation Metrics}
\label{s:evaluation-metrics}
Evaluating guitar tablatures requires domain-specific metrics, as conventional machine learning and NLP metrics miss crucial aspects of musicality and playability. Since there are no established standards in this field, we propose three metrics that evaluate both the accuracy of the transcription and the playability of the generated tablatures.

Preserving the original pitch is essential to maintain the musical integrity of a piece. The \textit{pitch accuracy} metric, ranging from 0\% to 100\%, measures how well the model reproduces the original pitches from the MIDI input. It allows for alternative string-fret combinations as long as the pitch remains correct.

The \textit{tab accuracy}, also ranging from 0\% to 100\%, reflects how well the professionally created ground-truth tablatures agree with the estimated fretting. This metric compares the predicted string-fret combinations with the ground truth, which is assumed to represent the optimal playing positions. This metric reflects the overall playability and consistency with the original piece.

A modified version of the difficulty estimation framework \cite{Radicioni.2005.Guitar} is used to objectively evaluate the playability of tablatures. The scoring system takes into account two types of movement: horizontal shifts along the fretboard (\textit{along}) and vertical shifts across the strings (\textit{across}). The difficulty of transitioning between two positions $(p, q)$ is calculated as follows: 

\begin{equation}
    \textit{difficulty}_{(p,q)} = \textit{along}_{(p,q)} + \textit{across}_{(p,q)} \quad,
\end{equation}
where
\begin{align}
    \textit{along}_{(p,q)} &= \textit{fret\_stretch}_{(p,q)} + \textit{locality}_{(p,q)} \quad,\\
    \textit{across}_{(p,q)} &= \textit{vertical\_stretch}_{(p,q)} \quad.
\end{align}

The $\textit{fret\_stretch}_{(p,q)}$ value measures the difficulty of the horizontal movement by calculating a delta between the frets of the first $\textit{p}$ and second position $\textit{q}$. Positive deltas, corresponding to movement to higher frets, are assumed to be easier as the fret spacing becomes shorter. Let $\Delta_\textit{fret} = q - p$, where $\textit{p, q}$ are the fret numbers. Then $\textit{fret\_stretch}_{(p,q)}$ is defined as: 

\begin{align}
\textit{fret\_stretch}_{(p,q)}  =
\begin{cases} 
0.50 \cdot |\Delta_{\textit{fret}}| & \text{if } \Delta_{\textit{fret}} > 0, \\
0.75 \cdot |\Delta_{\textit{fret}}| & \text{if } \Delta_{\textit{fret}} \leq 0 \quad.
\end{cases}
\end{align}

It also takes into account the location of the two positions, because the higher the fret, the more the string lifts off the fret and the harder it is to press the string. Locality is defined as
\begin{equation}
    \textit{locality}_{(p,q)} = \alpha \cdot (\textit{p} + \textit{q}) \quad, 
\end{equation}
where $\textit{p, q}$ are the fret numbers and $\alpha$ is a factor that should take into account the player’s
technical ability and the height of the strings. In the original article it was set at 0.25 by tests. 
The $\textit{vertical\_stretch}_{(p,q)}$ depends on the distance between the positions of the fingers on the adjacent strings. Since this depends strongly on which position is played with which finger, only standard values are used here. It is assumed that a delta of at most 1 is still comfortable when comparing the strings, so let $\Delta_\textit{string} = q - p$, where $\textit{p, q}$ are the string numbers. Then the $\textit{vertical\_stretch}_{(p,q)}$ is defined as:

\begin{align}
\textit{vertical\_stretch}_{(p,q)} =
\begin{cases} 
0.25 & \text{if } \Delta_{\textit{string}} \leq 1, \\
0.50 & \text{if } \Delta_{\textit{string}} > 1 \quad.
\end{cases}
\end{align}

To calculate the overall difficulty score for a tablature, we calculate the mean difficulty over all positions. The range is from 0 to 18.5 for a 24-fret guitar. The lowest difficulty, 0, is when notes are played on the same open string. The highest value is reached when jumping from the lowest open string to the highest string at the highest fret.

%% file: sections/experiments.tex
In this section, the results of our proposed model are presented in an evaluation on the test split of the GuitarToday, Leduc and DadaGP datasets.
For evaluation, the metrics described in Section \ref{s:evaluation-metrics} are used. 

\subsection{Data Encodings}
\label{s:data-encodings}
To evaluate the effects of data encoding on the transcription performance of guitar tablatures, we conducted an experiment with five different encoding strategies. Each encoding was designed to capture the essential musical information, and we experimented with different levels of abstraction and granularity. Table \ref{t:encodings} shows one note each as input and output text encoding.

\begin{table*}[t]
    \centering
    \begin{tabular}{  l  l  l  }
        \toprule
        \textbf{ID} & \textbf{Input} & \textbf{Target}\\ 
        \midrule
        \textit{v1} & NOTE\_ON$<$55$>$ & STRING$<$3$>$ FRET$<$0$>$ \\ 
        
        \textit{v2} & NOTE\_ON$<$55$>$ & TAB$<$3,0$>$ \\  
        
        \textit{v3} & NOTE\_ON$<$55$>$ TIME\_SHIFT$<$120$>$ NOTE\_OFF$<$55$>$ & TAB$<$3,0$>$ TIME\_SHIFT$<$120$>$ \\ 
        
        \textit{v4} & NOTE\_ON$<$55$>$ TIME\_SHIFT$<$120$>$ NOTE\_OFF$<$55$>$ & STRING$<$3$>$ FRET$<$0$>$ TIME\_SHIFT$<$120$>$ \\ 
        
        \textit{v5} & NOTE\_ON$<$55$>$ TIME\_SHIFT$<$120$>$ & TAB$<$3,0$>$ TIME\_SHIFT$<$120$>$ \\ 
        \bottomrule
    \end{tabular}
    \caption{Examples of the different input and target MIDI-to-text encodings based on the representation of one note. NOTE\_ON / NOTE\_OFF define the pitch and TIME\_SHIFT the duration. STRING and FRET, as well as the combined TAB token, define the corresponding string and fret for the pitch.}
    \label{t:encodings}
\end{table*}

The \textit{v1} encoding reduces the information to pitch, and the output shows the string and fret positions as separate tokens. By simplifying the input and focusing on pitch alone, this version tests whether pitch is sufficient without explicit timing cues. The \textit{v2} encoding further simplifies the \textit{v1} version by combining the string and fret information into a single \textit{TAB} token in the output. This further reduces the space required for the output tokens. The \textit{v3} encoding uses an event-based approach, using note-on and note-off events together with timeshift tokens to represent timing. The output also uses combined \textit{TAB} tokens for string and fret information, based on a compact representation of the tablature. This version includes both timing and pitch, allowing for a more comprehensive contextual understanding. Similar to \textit{v3}, the \textit{v4} encoding also uses the event-based approach, again separating the string and fret data into individual tokens in the output. The fifth encoding, \textit{v5}, also builds on \textit{v3}, but removes note-off tokens from the input, simplifies the event sequence and relies only on note-on and time-shift events to represent the duration of the note. The output uses combined \textit{TAB} tokens for string and fret information. This version explores the importance of note-off tokens in conveying time and duration information.

\begin{figure}[tb]
  \includegraphics[width=\columnwidth]{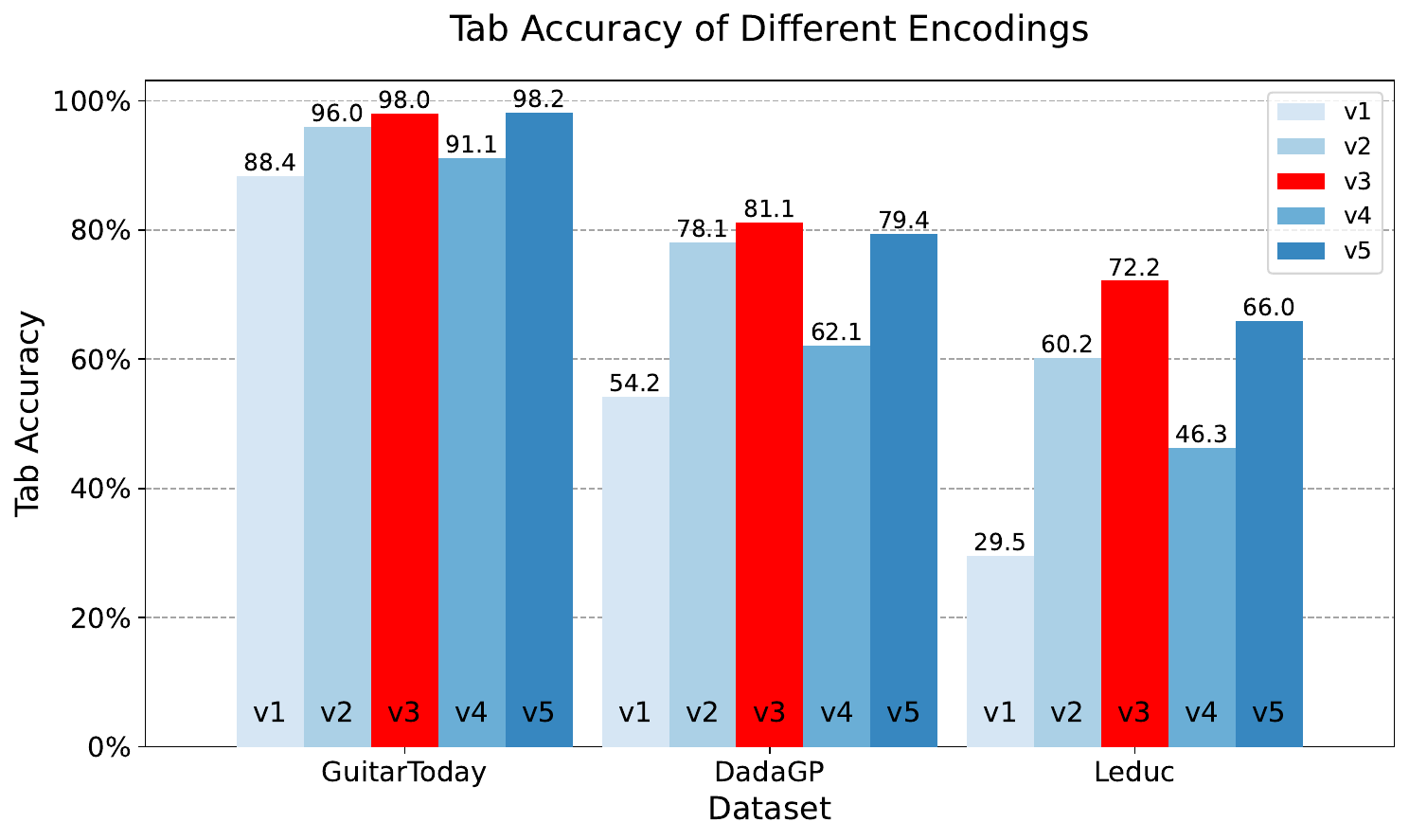}
  \centering
  \caption{Comparison of five different encodings across GuitarToday, DadaGP, and Leduc datasets by testing time and tab token variations. The combination of \textit{TAB} and \textit{TIME\_SHIFT} tokens in the \textit{v3} encoding showed the best result across the datasets.}
  \label{fig:encodings}
\end{figure}

The encodings were tested using the GuitarToday, DadaGP, and Leduc datasets and no post-processing was applied. The results in Figure \ref{fig:encodings} show varying levels of accuracy across the different encodings, suggesting that the choice of encoding has a significant influence on the model's performance. 

% \begin{figure}[tb]
%   \includegraphics[width=\columnwidth]{figures/bar_plot_encodings.pdf}
%   \centering
%   \caption{Comparison of different encodings across GuitarToday, DadaGP, and Leduc datasets}
%   \label{fig:encodings}
% \end{figure}

Both encoding \textit{v1} and \textit{v4} show that the division into \textit{STRING} and \textit{FRET} tokens is less accurate than the combined \textit{TAB} token. This is because when a pitch is correctly mapped, the probability of selecting one correct token is easier than selecting two correct tokens. The addition of timeshift information (in \textit{v3}, \textit{v4}, and \textit{v5}) improves tab accuracy, suggesting that the model benefits from explicit time and duration data. The reduction of tokens in \textit{v5} seems to remove context from the model. A comparison of the datasets shows that the combination of \textit{TAB} and \textit{TIME\_SHIFT} in the \textit{v3} encoding is well generalizable across different styles, which can be further improved by post-processing corrections.

\subsection{Effects of Post-Processing}
\label{s:results_post}
Post-processing plays a critical role in refining the output of the Fretting-Transformer model by addressing residual inaccuracies in the generated tablatures. The initial model outputs, while largely accurate in pitch, occasionally produce string-fret combinations that result in incorrect pitches or implausible fingerings. To mitigate these issues, two post-processing methods were implemented: overlap correction and neighbor search (see Section \ref{s:post-processing}).

The results of applying these methods are presented in Table \ref{tab:post_processing}. Without post-processing, the model achieves a pitch accuracy of 97.23\% and a tab accuracy of 68.56\%. Introducing overlap correction improves these metrics to 99.92\% and 72.15\%, respectively. Adding neighbor search further refines the output, resulting in perfect pitch accuracy (100.00\%) and a slight increase in tab accuracy to 72.19\%. These results highlight the importance of post-processing in enhancing the playability and musical fidelity of the generated tablatures, ensuring that the transcriptions remain both accurate and practical for guitarists.

\input{tables/post_processing.tex}

\subsection{Domain Adaption from Text}
\label{s:results_domain}

Pre-trained models in different sizes are available for the T5 transformer. These pre-trained models have been trained unsupervisedly on a large corpus of English texts. In this experiment, the pre-trained \textit{t5-small} is trained via domain adaption on the task of transcribing MIDI to tablature.
We compare the progression of the validation loss during training. 
Although the \textit{t5-small} configuration is larger than our proposed custom model, the training converges significantly faster. The choice of the optimizer also has a big impact. Choosing the Adafactor optimizer over AdamW improves the convergence speed of the custom model. 
Besides the faster convergence, the custom model optimized using Adafactor achieves the best tab accuracy, resulting in a $4\%$ increase in comparison to the pre-trained \textit{t5-small} model optimized with the same optimizer. 

% \begin{figure}[tb]
%   \includesvg[width=\columnwidth]{figures/exp6_val_loss.svg}
%   \centering
%   \caption{Training progress comparison between using a text pre-trained model and training from scratch}
%   \label{fig:domain_adaption}
% \end{figure}

\subsection{Alternative NLP Task Formulations}
\label{s:results_nlp}

\begin{table*}[t]
\centering
\begin{tabular}{l|cc|cc|cc}
\toprule
\textbf{Model} & \multicolumn{2}{c|}{\textbf{GuitarToday}} & \multicolumn{2}{c|}{\textbf{Leduc}} & \multicolumn{2}{c}{\textbf{DadaGP}} \\
\cmidrule{2-7}
& \textbf{Tab Accuracy} & \textbf{Difficulty} & \textbf{Tab Accuracy} & \textbf{Difficulty} & \textbf{Tab Accuracy} & \textbf{Difficulty} \\
\midrule
T5   & \textbf{98.41\%} & \textbf{1.9545} & \textbf{72.19\%} & \textbf{4.2415} & \textbf{81.58\%} & \textbf{2.4122} \\
BERT & 96.71\% & 2.0680 & 69.89\% & 4.4793 & 78.45\% & 2.7118 \\
GPT2 & 97.70\% & 1.9951 & 71.55\% & 4.2778 & 81.10\% & 2.4909 \\
\bottomrule
\end{tabular}
\caption{Comparison of models on GuitarToday, Leduc, and DadaGP test datasets (post-processing applied). The highest tab accuracy and lowest difficulty scores are printed in bold. Overall, all models achieve good results, with the T5 outperforming the others with a slightly higher tab accuracy and lower difficulty scores for all datasets.}
\label{tab:nlp_taks}
\end{table*}
The Fretting-Transformer interprets the task of transcribing MIDI to tablature as a translation between the MIDI language and the tablature language. Alternatively, the fill-mask and the text completion interpretation can also be applied. Therefore, two alternative architectures are explored.

The BERT architecture \cite{Devlin.2019}, with a configuration similar to that described in \cite{Edwards.2024}, is trained using masked language modeling. Here, the \textit{TAB$<$\#,\#$>$} tokens directly follow the \textit{NOTE\_ON$<$\#$>$} tokens in the input and are replaced by a \textit{$<$MASK$>$} token during training.

For implementing the text completion task interpretation, the GPT2 model \cite{Radford.2019.Language} is used. The model is trained to generate tablature tokens for a given sequence of MIDI tokens in a text completion manner. Therefore, a `MIDI:' token followed by the MIDI notes and a `TABS:' token is used as a primer sequence. The GPT2 model now completes the text by adding the tab token sequence.

The results of the different interpretations of the NLP task can be compared in Table \ref{tab:nlp_taks}. Although each model is able to preserve the original pitch of a note, the T5 model leads to slightly higher tab accuracy and lower difficulty scores for all datasets.

\subsection{Conditioning on Tuning and Capo}
\label{s:results_conditioning}
% 3. tuning/capo conditioned
\begin{table*}[t]
\centering
\begin{tabular}{l|cc|cc|cc}
\toprule
\textbf{Model} & \multicolumn{2}{c|}{\textbf{GuitarToday}} & \multicolumn{2}{c|}{\textbf{Leduc}} & \multicolumn{2}{c}{\textbf{DadaGP}} \\
\cmidrule{2-7}
& \textbf{Tab Accuracy} & \textbf{Difficulty} & \textbf{Tab Accuracy} & \textbf{Difficulty} & \textbf{Tab Accuracy} & \textbf{Difficulty} \\
\midrule
T5   & \textbf{98.19\%} & \textbf{1.9838} & \textbf{73.02\%} & \textbf{4.1875} & 79.89\% & 2.5774 \\
BERT & 96.57\% & 2.0772 & 70.52\% & 4.4846 & 78.97\% & 2.7191 \\
GPT2 & 97.61\% & 1.9995 & 70.60\% & 4.3209 & \textbf{81.45\%} & \textbf{2.4908} \\
\bottomrule
\end{tabular}
\caption{Comparison of tuning and capo conditioned models on GuitarToday, Leduc, and DadaGP test datasets (post-processing applied). The highest tab accuracy and lowest difficulty values are printed in bold. The T5 model outperforms the others in the higher quality GuitarToday and Leduc datasets. For DadaGP, GPT2 is slightly closer to the ground truth. }
\label{tab:tuning_capo_conditioning}
\end{table*}

In the previous experiments, the tablature transcription was examined for the case of standard tuning and without the usage of a capo. While this is true for the majority of tablatures available online, in some cases guitarists prefer to use alternative tunings or want to transpose the piece to a different key using a capo.
To incorporate these additional conditions, additional tokens are added for the tuning and the fret to which the capo is set.
The results of the tuning and capo conditioning can be seen in Table \ref{tab:tuning_capo_conditioning}. For a quantitative evaluation, the test splits of the aforementioned datasets were augmented in advance with random variations in tuning and capo usage.
For the GPT2 variant the tuning and capo conditions are introduced after the `TABS:' token in the primer sequence. By removing the conditioning tokens from the primer, it is also possible to let the model suggest a suitable tuning and capo option.

The results show that the generation of high quality tablatures also works very well on a conditional basis. Again, the T5 model outperforms the others in the higher quality GuitarToday and Leduc datasets. For DadaGP, GPT2 is slightly closer to the ground truth.

\subsection{Comparison with Baselines}
\label{s:results_overview}
% 1. State of art vergleich
% -dadagp, leduc, guitartoday
\begin{table*}[t]
\centering
\begin{tabular}{l|cc|cc|cc}
\toprule
\textbf{Method} & \multicolumn{2}{c|}{\textbf{GuitarToday}} & \multicolumn{2}{c|}{\textbf{Leduc}} & \multicolumn{2}{c}{\textbf{DadaGP}} \\
\cmidrule{2-7}
& \textbf{Tab Accuracy} & \textbf{Difficulty} & \textbf{Tab Accuracy} & \textbf{Difficulty} & \textbf{Tab Accuracy} & \textbf{Difficulty} \\
\midrule
\multicolumn{7}{c}{\textbf{Standard Tuning Without Capo}}\\
\midrule
Baseline & 98.30\% & \textbf{1.9362} & 58.11\% & 3.5365 & 79.21\% & 2.1082 \\
A$^*$   & 89.39\% & 2.1715 & 62.60\% & \textbf{2.5704} & 78.92\% & 3.8590 \\
TuxGuitar   & 98.30\% & \textbf{1.9362} & 57.94\% & 3.5327 & 78.83\% & \textbf{2.1043} \\
Guitar Pro   & 97.58\% & 1.9651 & 56.03\% & 4.0736 & 76.58\% & 2.2257 \\
Ours     & \textbf{98.41\%} & 1.9545 & \textbf{72.19\%} & 4.2415 & \textbf{81.58\%} & 2.4122 \\
\midrule
\multicolumn{7}{c}{\textbf{Capo and Tuning Conditioned (Standard, Half-Step, Full-Step, Drop-D)}}\\
\midrule
Baseline & 98.30\% & \textbf{1.9362} & 58.18\% & \textbf{3.5418} & 79.39\% & \textbf{2.1134} \\
A$^*$   & 94.89\% & 2.0359 & 62.26\% & 3.8686 & 79.17\% & 2.2191 \\
TuxGuitar   & 80.23\% & 2.5235 & 59.25\% & 3.7815 & 65.48\% & 2.5330 \\
Guitar Pro   & 97.58\% & 1.9651 & 56.17\% & 4.0756 & 76.13\% & 2.2930 \\
Ours     & \textbf{98.19\%} & 1.9838 & \textbf{73.02\%} & 4.1875 & \textbf{79.89\%} & 2.5774 \\
\bottomrule
\end{tabular}
\caption{Comparison of our approach (post-processing applied) with baseline and state-of-the-art methods across GuitarToday, Leduc, and DadaGP datasets under two experimental conditions: standard tuning without capo and capo/tuning-conditioned scenarios. The highest tab accuracy and lowest difficulty scores are printed in bold. Our proposed method achieves the highest tablature accuracy across all datasets in both scenarios, showing significant improvements and thereby surpassing both baseline and state-of-the-art approaches.}
\label{tab:sota_comparison}
\end{table*}

A comparison of the proposed Fretting-Transformer model with the baseline and state-of-the-art methods is shown in Table \ref{tab:sota_comparison}.
As a simple baseline, the string-fret combination with the lowest possible fret for a given pitch is chosen.
The A$^*$ algorithm method is a reimplementation of \cite{Burlet.2013} selected for comparison with the state of the art. 
All of the methods are able to provide valid tablatures for the given pitches and achieve $100\%$ pitch accuracy.
The Fretting-Transformer outperforms the A$^*$ and the baseline on all three datasets in terms of tab accuracy. Especially in the high quality tabs of the GuitarToday and the Leduc dataset, there is a significant difference. 
On DadaGP, the baseline already achieves quite good accuracies. This could confirm that a lot of the tablatures in this dataset are algorithmically generated. 
Besides the baseline methods, the commercial tool Guitar Pro 8.1.3\footnote{\url{https://guitar-pro.com}} and the open source tool TuxGuitar 1.6.6\footnote{\url{https://tuxguitar.app}} have been evaluated.

Our proposed method achieves the highest tablature accuracy across all datasets in both scenarios, showing significant improvements especially on Leduc and DadaGP, thereby surpassing both baseline and state-of-the-art approaches. This highlights the robustness of the Fretting-Transformer in handling a variety of musical styles and complexities, particularly with datasets like Leduc that feature intricate fretting tasks.
There exists a trade-off between tablature accuracy and difficulty. Methods such as Baseline and TuxGuitar achieve lower difficulty ratings, yet they fall short of the proposed method's accuracy. This indicates that an exclusive focus on playability may not reflect the preferences of guitarists and hence does not lead to optimal tablatures.

% 3.5 tuning und capo estiamtion with gpt2?

\subsection{Discussion}
\begin{figure}[tb]
  \includegraphics[width=\columnwidth]{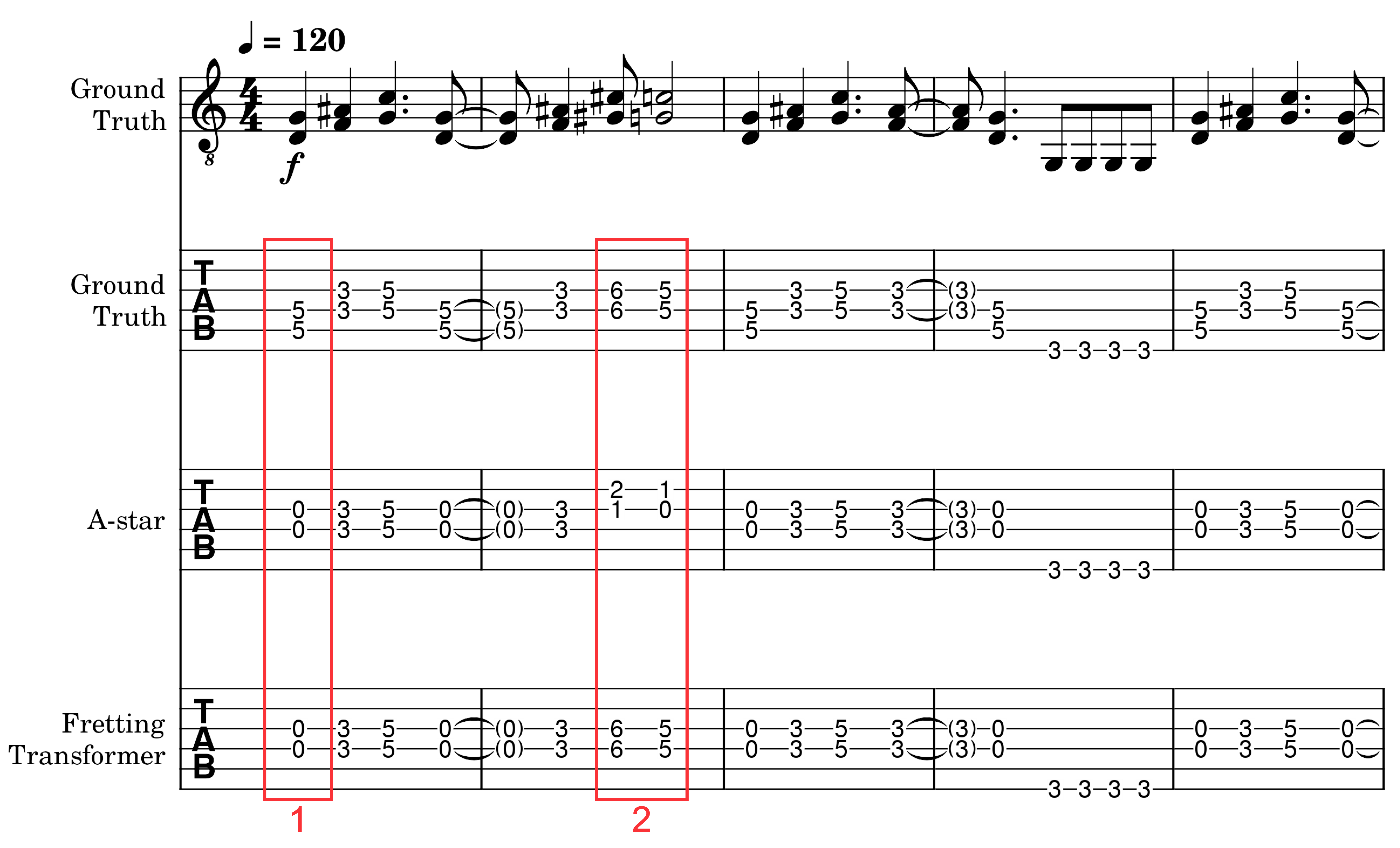}
  \centering
  \caption{Qualitative Comparison of ground truth, A$^*$ and Fretting-Transformer tablatures for `Smoke On The Water’. Box 1: Although the Fretting-Transformer varies from the ground truth, the use of the open strings might be preferred by many guitarists. Box 2: Our model is provided with more context and tends to make more consistent fretting decisions than A$^*$.}
  \label{fig:example}
\end{figure}
While achieving high tablature accuracy can be interpreted as how well the ground truth matches our predictions, the quality and the purpose of the tablatures provided can vary depending on the data source and the target audience. Figure \ref{fig:example} shows a short excerpt of the piece `Smoke on the Water' from the GuitarToday dataset.
Firstly, it can be noticed that although the Fretting-Transformer's result varies from the ground truth (see box 1), the use of the open strings might be preferred by many guitarists. This also highlights the limitations of the chosen metrics.
Secondly, when compared to the A$^*$ algorithm, the advantage of more contextual information becomes clear. While A$^*$ only looks at the previous and the next note, our model is provided with more context and tends to make more consistent fretting decisions (see box 2).

%% file: tables/post_processing.tex
\begin{table}[t]
  \centering
  \begin{tabular}{lccc}
    \toprule
    Post-Processing Method   & \textbf{Pitch Acc.} & \textbf{Tab Acc.}              \\
    \midrule
    No Post-Processing  & $\SI{97.23}{\percent}$          & $\SI{68.56}{\percent}$                \\
    Overlap & $\SI{99.92}{\percent}$          & $\SI{72.15}{\percent}$                \\
    Overlap + Neighbor Search & $\mathbf{\SI{100.00}{\percent}}$ & $\mathbf{\SI{72.19}{\percent}}$  \\
    \bottomrule
  \end{tabular}
  \caption{Evaluation of the effect of post-processing on the Leduc test dataset. With applying overlap and neighborhood search, perfect pitch accuracy can be achieved, which is necessary for musical fidelity.}
  \label{tab:post_processing}
\end{table}

%% file: sections/conclusion.tex
% conclusion
This research demonstrates the effectiveness of transformer models, particularly the T5 architecture, in transcribing MIDI sequences into guitar tablatures. By framing the task as a symbolic translation problem, the Fretting-Transformer outperforms existing methods, including A$^*$ and commercial tools like Guitar Pro. Key contributions include innovative data pre-processing, tokenization strategies, and conditioning on tuning and capo settings, which address challenges like string-fret ambiguity and playability.
The metrics presented proved to be helpful in assessing the overall quality of the tablatures and providing an assessment of playability.
The model’s adaptability to diverse datasets (DadaGP, GuitarToday, and Leduc) highlights its generalizability. Augmentation strategies and post-processing techniques, such as overlap correction, enhance accuracy and musical fidelity. However, limitations such as dataset constraints and the omission of advanced musical features like dynamics remain areas for future work.

Expanding datasets, incorporating additional musical features like voicing information, and combining this approach with other MIDI to score approaches could further enhance the model’s capabilities.